\documentclass[runningheads]{svmult}

\usepackage{makeidx}   
\usepackage{graphicx}  
\usepackage{subeqnar}  
\usepackage{multicol}  
\usepackage{physprbb}  
\makeindex             

\usepackage{amssymb}
\usepackage{amsmath}
\usepackage{epsfig}
\newcommand {\etal}{\begin{itshape}et al\end{itshape}.}
\newcommand {\Sr}{Sr$_2$RuO$_4$ }
\newcommand {\He}{$^3$He }

\begin{document}
\title*{The Behavior of a Triplet Superconductor in a Spin Only Magnetic Field }
\toctitle{The Behavior of a Triplet Superconductor in a Spin Only Magnetic
Field}

%
%
\titlerunning{The Behavior of a Triplet Superconductor
in a Spin Only Magnetic Field}
%
\author{B. J. Powell \and James F. Annett \and B. L. Gy\"{o}rffy}
\authorrunning{Powell et al.}
%
%
\institute{H. H. Wills Physics Laboratory, University of Bristol,
Tyndall Avenue, BS8 1TL, UK}

\maketitle              

\begin{abstract}
We investigate the order parameter of \Sr in an exchange-only
magnetic field. A Ginzburg-Landau symmetry analysis implies three
possibilities: a pure \He A phase, a \He A$_1$ or a \He A$_2$
phase. We explore the  exchange field dependence of the order
parameter and energy gap in a one-band model of Sr$_2$RuO$_4$. The
numerical solutions show no A$_1$ phase and that the A$_2$ phase
is lower in free energy than the A phase. We explore heat capacity
as a function of temperature and field strength and find
quantitatively different behaviors for the A and A$_2$ phases.
\end{abstract}

\section{Introduction}

A spin triplet superconductor should show a number of interesting
magnetic-field effects which are direct consequences of the
magnetic moment of the Cooper pairs. In particular, for
spin-triplet superconductors the Zeeman coupling between the
quasiparticle spins an external magnetic field need not lead to
Pauli limiting, unlike the case of spin-singlet superconductors.
In the extreme high field limit with completely exchange-split
bands we could expect single spin pairing of the majority spin
Fermi surface.  We may also expect possible phase transitions or
symmetry changes of the order parameter in a magnetic field, which
are analogous to the transitions seen in superfluid
$^3$He\cite{Lee,Vollhardt,Min}. The \He B-phase is destroyed in a
magnetic field in a qualitatively similar manner to a singlet
superconductor. On the other hand, if the zero field ground state
is one of equal-spin-pairing (ESP), then the gap function can
deform continuously as a function of a Zeeman field the \He
A-phase evolves first smoothly into A$_2$ phase and then, via a
phase transition A$_1$ phase, as it progresses from
equal-spin-pairing to single spin pairing with increasing
field\cite{Brussaard}.

The superconductor \Sr should be an ideal candidate to examine
these effects. There is strong evidence for spin-triplet
pairing\cite{Maeno} from direct measurements of
spin-susceptibility in the superconducting
state\cite{Ishida,Duffy}.  It is has a simple and well understood
Fermi surface\cite{FS}, and  is in the clean limit.  The detailed
gap function is still somewhat controversial, but is generally
believed\cite{Agterberg} to be  of tetragonal $E_u$
symmetry\cite{Annett}, and more specifically to be a
two-dimensional analogue of the \He A-phase, with $ {\bf d}({\bf
k}) \sim (\sin{k_x} + i \sin{k_y})(0,0,1)$. This order parameter
would agree with the spin-susceptibility measurements in the
superconducting state\cite{Ishida,Duffy}, and also would lead to
time-reversal symmetry breaking below $T_c$\cite{Luke}. More
recently,  specific heat\cite{NishiZaki}, penetration
depth\cite{DalevH} and thermal conductivity\cite{izawa}
experiments have shown that the gap must have line-nodes on the
Fermi surface. However, for the cylindrical Fermi surface
geometry\cite{FS} of \Sr a complete group theoretic analysis of
symmetry distinct pairing states does not show any which both
break time-reversal symmetry and have line-nodes\cite{Annett}. A
possible resolution to this dilemma has been developed in the
orbital dependent pairing model of Zhitomirsky and
Rice\cite{Zhitomirsky}
 and a in related model by
Litak \etal\cite{Litak}. For different reasons both groups
proposed that the gap function is of the form
\begin{equation}
 {\bf d}({\bf k}) \sim
(\sin{k_x} + \I\sin{k_y})(0,0,1)
\end{equation}
 on the dominant $\gamma$-Fermi surface sheet, and of the form
\begin{equation}
 {\bf d}({\bf k}) \sim
\left(
\sin{\left(\frac{k_x}{2}\right)}\cos{\left(\frac{k_y}{2}\right)} +
i \sin{\left(\frac{k_y}{2}\right)}\cos{\left(\frac{k_x}{2}\right)}
\right) \cos{\left(\frac{c k_z}{2}\right)}(0,0,1)
\end{equation}
on the $\alpha$ and $\beta$ sheets. Both of these functions
possess the same $E_u$ symmetry, but correspond to intra-plane and
inter-plane pairing interactions respectively. This gap function
has horizontal line nodes at $k_z=\pm \pi/c$ on the $\alpha$ and
$\beta$ sheets, and was shown to be in good agreement with
experimental temperature dependences for specific heat,
penetration depth and thermal conductivity\cite{Litak}.

In a magnetic field \Sr shows a number of
 unusual features. Firstly the vortex lattice is square\cite{Riseman,Kealey}
which agrees well with the predictions of a two-component $E_u$
symmetry Ginzburg-Landau theory\cite{Agterberg98,Heeb}.  Secondly
there is an anomalous second feature close to  $H_{c2}$, which
only occurs when the field is aligned within $1^o$   of the a-b
plane\cite{NishiZaki}. At the present time the origin of this
feature is uncertain. It may be a vortex lattice phase transition,
or it may correspond to a change in pairing symmetry with field,
perhaps analogous to the double superconducting transition in
UPt$_3$\cite{upt3}.

In this paper we will focus specifically on the unique effects of
the Cooper pair spin in a triplet superconductor. Therefore we
neglect the effects of the vector potential on the quasiparticles,
and instead focus solely on the Zeeman coupling of the
quasiparticle spin to the magnetic field. We can justify this
model by appealing to the strong Stoner enhancement in
Sr$_2$RuO$_4$\cite{Discovery,M&S}, and so the exchange field will
be large. Alternatively, our model may be appropriate for  the
ferromagnetic superconductor ZrZn$_2$\cite{Pfleiderer}.

This paper is organised as follows. Firstly we write a simple
single-band model Hamiltonian for $p$-wave pairing  in the
$\gamma$-band of \Sr. Next we examine how a spin-only magnetic
field enters the  corresponding $E_u$  symmetry Ginzburg-Landau
theory.  In section 4,  we present detailed numerical results for
the field dependent energy gap, and specific heat for the two
relevant cases of the exchange field  either parallel or
perpendicular to the ${\bf d}({\bf k})$ order parameter. We show
that the lower free energy state in analogous to the \He A$_2$
phase. Finally, in Sec. 5 we present our conclusions.

\section{A Microscopic Model for a Triplet Superconductor
in a Spin Only Magnetic Field}

We consider the effect of a spin-only magnetic field, $\vec{H}$,
on an attractive, nearest neighbour, Hubbard model. We use a
one-band model, appropriate for the $\gamma$ sheet of the
Sr$_2$RuO$_4$ Fermi surface. The set of interaction constants,
$U_{ij}^{\sigma\sigma'},$ describe attractions between electrons
on sites $i$ and $j$ with spins $\sigma$ and $\sigma'$. The
Hamiltonian for this model is:

\begin{equation}
\hat{\cal{H}}=\sum_{ij\sigma}((\epsilon-\mu)\delta_{ij}-t_{ij})\hat{c}_{i\sigma}^\dag\hat{c}_{j\sigma}
-\frac{1}{2}\sum_{ij\sigma\sigma'}U_{ij}^{\sigma\sigma'}\hat{n}_{i\sigma}\hat{n}_{j\sigma}
+\mu_{B}\sum_{i\sigma\sigma'}\hat{c}_{i\sigma}^\dag(\vec{\sigma}_{\sigma\sigma'}\cdot\vec{H})\hat{c}_{i\sigma}
\end{equation}
where $t_{ij}$ is the hopping integral, $\epsilon$ is the site
energy, $\hat{c}_{i\sigma}^{(\dag)}$ are the usual annihilation
(creation) operators and $\hat{n}_{i\sigma}$ is the number
operator. $\vec{\sigma}_{\sigma\sigma'}$ are the components of the
vector of Pauli matrices:
\begin{equation}
\underline{\underline{\vec{\sigma}}}=(\underline{\underline{\sigma_1}},
\underline{\underline{\sigma_2}},\underline{\underline{\sigma_3}})
\end{equation}
By making the Hartree-Fock-Gorkov approximation and taking a
lattice Fourier transform the following spin-generalised
Bogoliubov-de Gennes (BdG) equation can be derived from the above
Hamiltonian.

\begin{eqnarray}
\begin{pmatrix}
\epsilon _{\vec k}+\mu_BH_3&\mu_B(H_1-\I H_2)&\Delta
_{\uparrow\uparrow }({\vec{k}})&\Delta_{\uparrow\downarrow
}(\vec{k})
\\ \mu_B(H_1+\I H_2)&\epsilon _{\vec{k}}-\mu_BH_3 &\Delta_{\downarrow\uparrow }({\vec{k}}) &\Delta_{\downarrow\downarrow }({\vec{k}})
\\ -\Delta _{\uparrow\uparrow }^\dagger (-\vec{k})&-\Delta_{\uparrow\downarrow }^\dagger (-\vec{k})&-\epsilon_{-\vec{k}}-\mu_BH_3&\mu_B(-H_1-\I H_2)
\\ -\Delta _{\downarrow\uparrow }^\dagger (-\vec{k})&-\Delta_{\downarrow\downarrow }^\dagger
(-\vec{k})&\mu_B(-H_1+\I H_2)&-\epsilon_{-\vec{k}}+\mu_BH_3
\end{pmatrix}
\begin{pmatrix}
u_{\uparrow\sigma}({\vec{k}}) \\ u_{\downarrow\sigma}({\vec{k}}) \\
v_{\uparrow\sigma}({\vec{k}}) \\ v_{\downarrow\sigma}({\vec{k}})
\end{pmatrix} \notag
\\ \hspace{85mm}
=E_{\sigma}({\vec{k}})
\begin{pmatrix}
u_{\uparrow\sigma}({\vec{k}}) \\ u_{\downarrow\sigma}({\vec{k}}) \\
v_{\uparrow\sigma}({\vec{k}}) \\ v_{\downarrow\sigma}({\vec{k}})
\end{pmatrix},
\end{eqnarray}
where $\epsilon_{\vec{k}}$ is the (Fourier transformed) normal,
spin independent part of the Hamiltonian. The order parameters
$\Delta _{\sigma\sigma'}({\vec{k}})$ are determined self
consistently by

\begin{equation}
\label{eqn:self_const} \Delta_{\sigma\sigma'}({\bf
k})=-\frac{1}{2}\sum_{\vec{q}\sigma''}U^{\sigma\sigma'}(\vec{q})
(u_{\sigma\sigma''}(-\vec{q})v^\ast_{\sigma'\sigma''}(-\vec{q})-
v^\ast_{\sigma\sigma''}(\vec{q})u_{\sigma'\sigma''}(\vec{q}))(1-2f_{\vec{q}\sigma''})
\end{equation}
where $U^{\sigma\sigma'}(\vec{q})$ is the lattice Fourier
transform of $U_{ij}^{\sigma\sigma'}$ and $f_{\vec{q}\sigma}$ is
shorthand for the Fermi function $f(E_{\vec{q}\sigma})$. It is
natural to separate the spin-generalized BdG equation into triplet
and singlet parts:

\begin{equation}
\underline{\underline{\Delta}}({\vec{k}})\equiv
\begin{pmatrix}
\Delta
_{\uparrow\uparrow}(\vec{k})&\Delta_{\uparrow\downarrow}({\vec{k}})
\\ \Delta_{\downarrow\uparrow }({\vec{k}}) &\Delta_{\downarrow\downarrow }({\vec{k}})
\end{pmatrix}
=(d_0(\vec{k})+\underline{\underline{\sigma}}\cdot\vec{d}(\vec{k}))\I\underline{\underline{\sigma_2}}.
\end{equation}
$d_0(\vec{k})$ is the (scalar) singlet order parameter and
$\vec{d}(\vec{k})$ is the (vector) triplet order parameter. The
singlet order parameter is symmetric under spatial inversion while
the triplet order parameter in anti-symmetric under spatial
inversion. Hence, the BdG equation can be rewritten as

\begin{eqnarray}
\begin{pmatrix}
\epsilon _{\vec{k}}+\mu_BH_3 & \mu_B(H_1-\I H_2) &
-d_{1}(\vec{k})+\I d_{2}({\vec{k}}) &
d_{0}({\vec{k}})+d_{3}({\vec{k}})
\\ \mu_B(H_1+\I H_2)&\epsilon _{\vec{k}}-\mu_BH_3 & -d_{0}({\vec{k}})+d_{3}({\vec{k}}) & d_{1}({\vec{k}})+\I d_{2}({\vec{k}})
\\ -d^\ast_{1}({\vec{k}})-d^\ast_{2}({\vec{k}}) & -d^\ast_{0}({\vec{k}})+d^\ast_{3}({\vec{k}}) &-\epsilon_{\vec{k}}-\mu_BH_3&\mu_B(-H_1-\I H_2)
\\ d^\ast_{0}({\vec{k}})+d^\ast_{3}({\vec{k}}) & d^\ast_{1}({\vec{k}})-d^\ast_{2}({\vec{k}})
&\mu_B(-H_1+\I H_2)&-\epsilon_{\vec{k}}+\mu_BH_3
\end{pmatrix}
\begin{pmatrix}
u_{\uparrow\sigma}({\vec{k}}) \\ u_{\downarrow\sigma}({\vec{k}}) \\
v_{\uparrow\sigma}({\vec{k}}) \\ v_{\downarrow\sigma}({\vec{k}})
\end{pmatrix} \notag
\end{eqnarray}
\begin{eqnarray}
\hspace{80mm} =E_{\sigma}({\vec{k}})
\begin{pmatrix}
u_{\uparrow\sigma}({\vec{k}}) \\ u_{\downarrow\sigma}({\vec{k}}) \\
v_{\uparrow\sigma}({\vec{k}}) \\ v_{\downarrow\sigma}({\vec{k}})
\end{pmatrix}.
\end{eqnarray}

If there is no superconductivity in the triplet channel we regain
the standard result \cite{Min} for the spectrum of a singlet
superconductor in a spin only magnetic field:

\begin{equation}
E(\vec{k})=\pm\sqrt{\epsilon_{\vec{k}}^2+|d_0(\vec{k})|^2}\pm\mu_B|\vec{H}|.
\end{equation}
By setting the singlet order parameter to zero we find that the
equivalent result for a triplet superconductor is

\begin{equation}
E(\vec{k})=\pm\sqrt{\epsilon_{\vec{k}}^2+\mu_B^2|\vec{H}|^2+|\vec{d}(\vec{k})|^2\pm\sqrt{\Lambda(\vec{k})}}
\end{equation}
where
\begin{eqnarray}
\Lambda(\vec{k}) &=&
|\vec{d}(\vec{k})\times\vec{d}(\vec{k})^\ast|^2 +
4\epsilon_{\vec{k}}^2\mu_B^2|\vec{H}|^2 +
4\mu_B^2|\vec{H}\cdot\vec{d}(\vec{k})|^2 \notag
\\ && +
4\I\epsilon_{\vec{k}}\mu_B\vec{H}\cdot\vec{d}(\vec{k})\times\vec{d}(\vec{k})^\ast.
\label{spectrum}
\end{eqnarray}
It should be noted that this does not assume a unitary order
parameter\footnote{A unitary state is any state for which
$\vec{d}(\vec{k})\times\vec{d}^\ast(\vec{k})=0$.}.

It is a relatively straightforward process to calculate
thermodynamic properties for a triplet superconductor. For example
the specific heat is given by

\begin{eqnarray}
C_V &=& T\frac{\partial S}{\partial T}
\\ &=& -k_BT\frac{\partial}{\partial
T}\sum_{\vec{k}\sigma}(f_{\vec{k}\sigma}ln(f_{\vec{k}\sigma})+(1-f_{\vec{k}\sigma})ln(1-f_{\vec{k}\sigma}))
\\ &=&
\sum_{\vec{k}\sigma}f_{\vec{k}\sigma}(1-f_{\vec{k}\sigma})\left(\frac{E_\sigma(\vec{k})^2}{k_BT^2}+\frac{1}{k_BT}E_\sigma(\vec{k})\frac{d}{dT}E_\sigma(\vec{k})\right)
\\ &=& \sum_{\vec{k}\sigma}\frac{f_{\vec{k}\sigma}(1-f_{\vec{k}\sigma})}{k_BT^2}\left(E_\sigma(\vec{k})^2-\frac{T}{2}\frac{d}{dT}|\vec{d}(\vec{k})|^2\right)
\end{eqnarray}

\section{Ginzburg--Landau Theory of a Quasi--Two Dimensional Triplet
Superconductor in a Magnetic Field}

Before considering numerical solutions of the self consistent
Bogoliubov--de Gennes equations, we will examine the possible
results by deriving a Ginzburg--Landau theory from our microscopic
theory.

Consider a quasi--two dimensional system with two orbital degrees
of freedom (which we label x and y) and three spin degrees of
freedom (labelled 1, 2 and 3.) 
Hence, instead of the familiar 3 by 3 order parameter of \He this
system is described by the complex 2 by 3 matrix $A$, which is
related to the microscopic order parameter, $\vec{d}(\vec{k})$, by

\begin{eqnarray}
\begin{pmatrix}
d_1(\vec{k})
\\ d_2(\vec{k})
\\ d_3(\vec{k})
\end{pmatrix}
=
\begin{pmatrix}
A_{1x} \sin{k_x} + A_{1y} \sin{k_y} \\ A_{2x} \sin{k_x} + A_{2y} \sin{k_y} \\
A_{3x} \sin{k_x} + A_{3y} \sin{k_y}
\end{pmatrix}
= \begin{pmatrix} \vec{A}_{x} & \vec{A}_{y}
\end{pmatrix}
\begin{pmatrix}
\sin{k_x} \\ \sin{k_y}
\end{pmatrix}.
\end{eqnarray}
In zero field the free energy for a tetragonal crystal is given
by\cite{Annett}

\begin{eqnarray}
\notag F &=& \alpha (T-T_c)(|\vec{A}_{x}|^2 + |\vec{A}_{y}|^2) +
\beta_1 (|\vec{A}_{x}|^2 + |\vec{A}_{y}|^2)^2  + \beta_2
|\vec{A}_{x} \cdot \vec{A}_{x} + \vec{A}_{y} \cdot \vec{A}_{y}|^2
\\ && \notag + \beta_3 ((\vec{A}_{x} \cdot \vec{A}_{y}^*)^2 + (\vec{A}_{x}^* \cdot \vec{A}_{y})^2
+ (\vec{A}_{x}^* \cdot \vec{A}_{x})^2 + (\vec{A}_{y}^* \cdot
\vec{A}_{y})^2) \\ && \notag + \beta_4 (2|\vec{A}_{x} \cdot
\vec{A}_{y}^*|^2 + |\vec{A}_{x}|^4 + |\vec{A}_{y}|^4) \\ && \notag
+ \beta_5 (2|\vec{A}_{x} \cdot \vec{A}_{y}|^2 + |\vec{A}_{x} \cdot
\vec{A}_{x}|^2 + |\vec{A}_{y} \cdot \vec{A}_{y}|^2) \\ &&  +
\beta_6 (|\vec{A}_{x} \cdot \vec{A}_{x}|^2 + |\vec{A}_{y} \cdot
\vec{A}_{y}|^2) + \beta_7 (|\vec{A}_{x}|^4 + |\vec{A}_{y}|^4).
\label{eqn:zero_free}
\end{eqnarray}
Only the first five quartic terms ($\beta_1-\beta_5$) are required
to describe \He \cite{B&M}. The additional two terms here
($\beta_6$ and $\beta_7$) appear because the rotational symmetry
of the crystal is discrete, where as rotational symmetry is
continuous in the fluid. Gradient terms can also be
calculated\cite{Annett,Agterberg98,Heeb}, but we will not make use
of these here.

To second order in $A$ the free energy in a finite magnetic field,
$F_{\vec{H}}$, is

\begin{equation}
F_{\vec{H}} = \frac{1}{\beta} \sum_{\I\omega_n} \int \D\vec{k}
\hspace{2mm} tr \left(
\underline{\underline{G}}_0(\vec{k},\I\omega_n)
\underline{\underline{\Delta}} (\vec{k})
\underline{\underline{G}}_0^*(-\vec{k},\I\omega_n)
\underline{\underline{\Delta}}^\dag (-\vec{k}) \right),
\end{equation}
where,

\begin{equation}
\underline{\underline{G}}_0(\vec{k},\I\omega_n) = \left(
\I\omega_n +
\varepsilon_{\vec{k}}-\mu+\mu_B\vec{\sigma}\cdot\vec{H}
\right)^{-1}
\end{equation}
and $\omega_n$ are the Matsubara frequencies.

Thus to all orders in $\vec{H}$

\begin{eqnarray}
\notag F_{\vec{H}} &=& -\frac{1}{\beta} \sum_{\I\omega_n}
\textrm{tr} \int \D\vec{k} \hspace{2mm} \frac{(\I\omega_n -
\varepsilon_{\vec{k}} + \mu + \mu_B\vec{\sigma} \cdot \vec{H})
}{\left[ (\I\omega_n - \varepsilon_{\vec{k}} + \mu)^2 -
|\vec{H}|^2 \right] }
\\ && \times (\vec{\sigma} \cdot \vec{A}_x\sin{k_x} + \vec{\sigma} \cdot
\vec{A}_y\sin{k_y}) \sigma_2 \frac{(\I\omega_n +
\varepsilon_{\vec{k}} - \mu - \mu_B\vec{\sigma}^* \cdot \vec{H})
}{\left[ (\I\omega_n + \varepsilon_{\vec{k}} - \mu)^2 -
|\vec{H}|^2 \right]} \notag \\ && \times (\vec{\sigma}^* \cdot
\vec{A}_x^*\sin{k_x} + \vec{\sigma}^* \cdot
\vec{A}_y^*\sin{k_y})\sigma_2.
\end{eqnarray}
Hence,

\begin{equation}
F_{\vec{H}} = \vec{A}_x \underline{\underline{\chi}}_{xx}
\vec{A}_x^* + \vec{A}_x \underline{\underline{\chi}}_{xy}
\vec{A}_y^* + \vec{A}_y \underline{\underline{\chi}}_{yx}
\vec{A}_y^* + \vec{A}_y \underline{\underline{\chi}}_{yy}
\vec{A}_y^*,
\end{equation}
where,

\begin{eqnarray}
\chi_{ij}^{\alpha\beta} &=& -\frac{1}{\beta} \sum_{\I\omega_n}
\int \D\vec{k} \sin{k_i}\sin{k_j} \\ \notag && \times tr \left(
\frac{(\I\omega_n - \varepsilon_{\vec{k}} + \mu +
\mu_B\vec{\sigma} \cdot \vec{H}) \sigma_\alpha\sigma_2 (\I\omega_n
+ \varepsilon_{\vec{k}} - \mu - \mu_B\vec{\sigma}^* \cdot \vec{H})
\sigma_\beta^*\sigma_2 }{ \left[ (\I\omega_n -
\varepsilon_{\vec{k}} + \mu)^2 - |\vec{H}|^2 \right] \left[
(\I\omega_n + \varepsilon_{\vec{k}} - \mu)^2 - |\vec{H}|^2
\right]} \right). \label{Zeeman_terms}
\end{eqnarray}
By x--y symmetry $\chi_{xy} = \chi_{yx} = 0$. Some algebra then
leads to

\begin{eqnarray}
F_{\vec{H}} &=& (\alpha_0 + \alpha_2|\vec{H}|^2)(|\vec{A}_{x}|^2 +
|\vec{A}_{y}|^2) + \I\alpha_1\vec{H} \cdot (\vec{A}_{x} \times
\vec{A}_{x}^* + \vec{A}_{y} \times \vec{A}_{y}^*) \notag \\ && -
2\alpha_2 |\vec{H} \cdot \vec{A}|^2.
\end{eqnarray}
Where,

\begin{eqnarray}
\alpha_0 &=& \frac{2}{\beta}\sum_{\I\omega_n}\int \D\vec{k}
\frac{\sin^2{k_x} \left((\varepsilon_{\vec{k}} - \mu)^2 +
\omega_n^2\right)}{ \left[ (\I\omega_n - \varepsilon_{\vec{k}} +
\mu)^2 - |\vec{H}|^2 \right] \left[ (\I\omega_n +
\varepsilon_{\vec{k}} - \mu)^2 - |\vec{H}|^2 \right] },
\\ \alpha_1 &=& -\frac{4\mu_B}{\beta}\sum_{\I\omega_n}\int \D\vec{k}
\frac{\sin^2{k_x} (\varepsilon_{\vec{k}} - \mu)}{ \left[
(\I\omega_n - \varepsilon_{\vec{k}} + \mu)^2 - |\vec{H}|^2 \right]
\left[ (\I\omega_n + \varepsilon_{\vec{k}} - \mu)^2 - |\vec{H}|^2
\right] }
\end{eqnarray} and
\begin{eqnarray}
\alpha_2 &=& -\frac{2\mu_B^2}{\beta}\sum_{\I\omega_n}\int
\D\vec{k} \frac{\sin^2{k_x} }{ \left[ (\I\omega_n -
\varepsilon_{\vec{k}} + \mu)^2 - |\vec{H}|^2 \right] \left[
(\I\omega_n + \varepsilon_{\vec{k}} - \mu)^2 - |\vec{H}|^2 \right]
}.
\end{eqnarray}
Clearly $\alpha_0$ reduces to $\alpha$ (\ref{eqn:zero_free}) in
zero magnetic field, but the $\alpha_1$ and $\alpha_2$ terms do
not have an analogue in the zero field Ginzburg--Landau expansion.
It is interesting to note the similarity of these extra terms to
the change in the Hartree--Fock--Gorkov quasiparticle spectrum
caused by the magnetic field - (\ref{spectrum}). The cross product
of any complex vector with its complex conjugate is purely
imaginary\footnote{This can easily be confirmed. Consider the
cross product of the most general complex vector, $\vec{v}=(a+ib,
c+id, e+if)$. It is trivial to show that $\vec{v} \times \vec{v}^*
= -2i(cf-de, be-af, ad-bc)$.} so the square root of minus one
before the $\alpha_1$ term in the expression for the free energy
is to be expected.

As we have expanded in $A$ but not in $\vec{H}$ the above
expression for the free energy is valid for small gaps at all
field strengths. It is therefore valid close to $H_c$. But, note
that, since we assumed an exchange-only magnetic field we do not
consider the vortex lattice here. Agterberg and
Heeb\cite{Agterberg98,Heeb} have discussed the vortex lattice
using Ginzburg--Landau theory, but did not include the Zeeman
terms of (\ref{Zeeman_terms}).

In the Ginzburg--Landau formalism the superconducting phase
transition occurs when the quadratic terms go to zero. In a zero
field this condition is simply

\begin{equation}
\alpha (T-T_C) = 0.
\end{equation}
In a finite spin only magnetic field the equivalent condition is
that the matrix

\begin{equation}
\underline{\underline{\alpha}} = \alpha_{ij}A_iA_j^*
\end{equation}
has (at least) one zero eigenvalue, but no negative eigenvalues,
where the indices $i$ and $j$ run over both orbital and spin
degrees of freedom. In this case

\begin{equation}
\underline{\underline{\alpha}} =
\begin{pmatrix}
\underline{\underline{\beta}} & 0 \\ 0 &
\underline{\underline{\beta}}
\end{pmatrix},
\end{equation}
where,
\begin{equation}
\underline{\underline{\beta}} =
\begin{pmatrix}
\alpha_0 + \alpha_2|\vec{H}|^2 - 2\alpha_2H_1^2 & \I\alpha_1H_3 &
-\I\alpha_1H_2 \\ -\I\alpha_1H_3 & \alpha_0 + \alpha_2|\vec{H}|^2
- 2\alpha_2H_2^2 & \I\alpha_1H_1  \\ \I\alpha_1H_2 &
-\I\alpha_1H_1 & \alpha_0 + \alpha_2|\vec{H}|^2 - 2\alpha_2H_3^2
\end{pmatrix}.
\end{equation}
The condition for there being a zero eigenvalue of
$\underline{\underline{\alpha}}$ is

\begin{eqnarray}
\left( \alpha_0 + \alpha_2|\vec{H}|^2 - 2\alpha_2H_1^2 \right)
\left( \alpha_0 + \alpha_2|\vec{H}|^2 - 2\alpha_2H_1^2 \right)
\left( \alpha_0 + \alpha_2|\vec{H}|^2 - 2\alpha_2H_1^2 \right) \notag \\
- \left( \alpha_0 + \alpha_2|\vec{H}|^2 - 2\alpha_2H_1^2 \right)
\alpha_1^2H_1^2  \left( \alpha_0 + \alpha_2|\vec{H}|^2 -
2\alpha_2H_2^2 \right) \alpha_1^2H_2^2 \notag \\ - \left( \alpha_0
+ \alpha_2|\vec{H}|^2 - 2\alpha_2H_3^2 \right) \alpha_1^2H_2^2 =
0.
\end{eqnarray}
This expression can be greatly simplified by choosing our
coordinate system so that $\vec{H}$ lies parallel to one of the
axes. With, for example, $\vec{H}=(0,0,H)$ we find

\begin{equation}
\underline{\underline{\beta}} =
\begin{pmatrix}
\alpha_0 + \alpha_2H^2 & \I\alpha_1H & 0 \\
-\I\alpha_1H & \alpha_0 + \alpha_2H^2 & 0 \\
0 & 0 & \alpha_0 - \alpha_2H^2
\end{pmatrix}.
\end{equation}
Which has at least one zero eigenvalue when

\begin{eqnarray}
(\alpha_0 - \alpha_2H^2)((\alpha_0+\alpha_2H^2)^2 - \alpha_1^2H^2
) = 0.
\end{eqnarray}
The eigenvectors of $\underline{\underline{\alpha}}$ are

\begin{eqnarray}
\begin{pmatrix}
A_{1x} \\ A_{2x} \\ A_{3x} \\ A_{1y} \\ A_{2y} \\ A_{3y}
\end{pmatrix} =
\begin{pmatrix}
0 \\ 0 \\ 1 \\ 0 \\ 0 \\ 0
\end{pmatrix},
\begin{pmatrix}
0 \\ 0 \\ 0 \\ 0 \\ 0 \\ 1
\end{pmatrix},
\begin{pmatrix}
1 \\ \I\kappa \\ 0 \\ 0 \\ 0 \\ 0
\end{pmatrix},
\begin{pmatrix}
0 \\ 0 \\ 0 \\ 1 \\ \I\kappa \\ 0
\end{pmatrix},
\begin{pmatrix}
-\I\kappa \\ 1 \\ 0 \\ 0 \\ 0 \\ 0
\end{pmatrix} \textrm{and}
\begin{pmatrix}
0 \\ 0 \\ 0 \\ -\I\kappa \\ 1 \\ 0
\end{pmatrix}.
\end{eqnarray}
Where $\kappa$ is real. To second order in $A$, $\kappa$ is given
by

\begin{equation}
\kappa = -\frac{\alpha_0+\alpha_2H^2}{\alpha_1H}
\end{equation}

Much recent work (see introduction) has suggested that \Sr is
likely to be in an state analogous to the A-phase of $^3$He. If
the pairing interaction favours the A-phase in zero magnetic field
there are three possible solutions in a magnetic field.

\begin{eqnarray}
\vec{A}_x=-\I\vec{A}_y=(0,0,1) \label{eqn:A-phase} \label{eqn:A_para_phase}\\
\vec{A}_x=-\I\vec{A}_y=(1, \I\kappa, 0) \label{eqn:A1-phasea} \\
\vec{A}_x=-\I\vec{A}_y=(-\I\kappa, 1, 0) \label{eqn:A1-phaseb}
\end{eqnarray}

Equation \ref{eqn:A-phase} is the A-phase with $\vec{d}(\vec{k})$
parallel to $\vec{H}$. Equations \ref{eqn:A1-phasea} and
\ref{eqn:A1-phaseb} both give the A$_2$-phase for $0<|\kappa|<1$
and the A$_1$-phase for $|\kappa|=1$. Analogy may be drawn to the
description of elliptically polarised light in
optics\cite{Born_Wolf}. One can think of the three A-like phases
as being described by an ellipse of eccentricity
$\sqrt{1-\kappa^2}$. The A-phase is the special case of linear
polarization when the ellipse reduces to a line parallel to
$\vec{d}(\vec{k})$. The A$_1$ phase is the special case of
circularly polarized light a circle which lies in the 1,2-plane.
The A$_2$ corresponds to any ellipse between these two extremes.
In the A$_1$ and A$_2$-phases by taking the appropriate
superposition of (\ref{eqn:A1-phasea}) and (\ref{eqn:A1-phaseb})
the major axis of the ellipse can be made to point in any
direction in the plane perpendicular to $\vec{H}$.

\section{Numerical results}

To progress further we must resort to solving the self consistent
Bogoliubov--de Gennes equations numerically. To do this we fit the
hopping integral and site energy to the experimentally determined
Fermi surface of the $\gamma$-sheet of \Sr  \cite{FS}. The
interaction potential is restricted to include nearest neighbour
terms only and chosen to give the experimentally observed critical
temperature (1.5K).

\subsection{$\vec{d}(\vec{k})$ parallel to $\vec{H}$}

We begin by studying the first solution of the Ginzburg--Landau
theory (\ref{eqn:A_para_phase}), in which $\vec{d}(\vec{k})$ is
parallel to $\vec{H}$. In zero field we find that the ground state
of the model is a triplet state analogous to the A-phase of
$^3$He, specifically the state is

\begin{equation}
\vec{d}=\Delta_{0}(\sin{k_{x}}+\I\sin{k_{y}})\vec{\hat{e}}.
\end{equation}
Here we have defined the vector order parameter to point in the
$\vec{e}$ direction. In zero field, all directions in spin space
are degenerate if spin--orbit coupling is neglected. When an
external field is applied the ground state has $\vec{d}(\vec{k})$
perpendicular to the field, as we will show below. However, in \Sr
the order parameter is thought to be aligned with the
c-axis\cite{Maeno}, by spin-orbit coupling. Therefore despite the
low critical field along the c-axis, \Sr presents us with the
possibility of studying a triplet superconductor with a magnetic
field parallel to the order parameter. It is therefore interesting
to predict what would be observed in such experiments. To do this
we simply discard any A-phase like solutions with
$\vec{d}(\vec{k})$ not parallel to $\vec{H}$. We then consider the
remaining self consistent solution of the BdG equation with the
lowest free energy.

\begin{figure}
    \centering
    \epsfig{figure=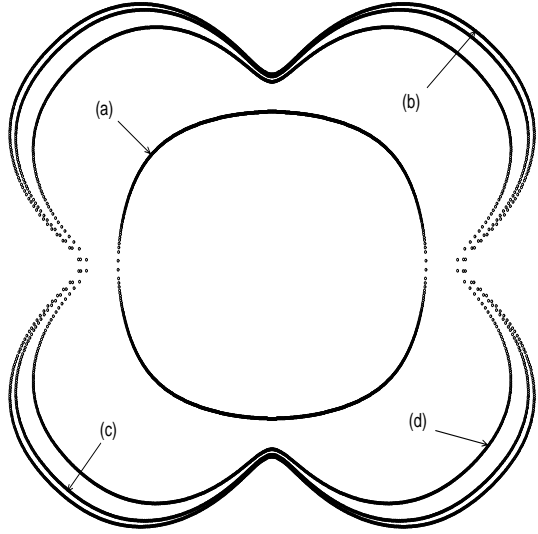, height=7cm, width=7cm, angle=270}
    \caption{(a)The Fermi surface and the gap at $T/T_C=0.5$ with (b) $\mu_BH/k_BT_C=0$, (c) $\mu_BH/k_BT_C=0.5$, (d) $\mu_BH/k_BT_C=0.9$}
\label{fig:para_gap}
\end{figure}

A field applied parallel to $\vec{d}(\vec{k})$ does not cause a
change in the symmetry of the gap. It follows that at zero
temperature the gap is independent of magnetic field strength (see
appendix). At finite temperature, a field applied parallel to the
order parameter causes a change in the magnitude of the gap (see
Fig. \ref{fig:para_gap}.) It should be noted that the gap is
nodeless but has minima at $k_x=0$ and $k_y=0$.

We calculate the heat capacity, magnetisation and magnetic
susceptibility as functions of temperature and field strength. For
an isotropic nodeless gap in zero field it is well known\cite{Min}
that these properties behave as
\begin{equation}
C_v,M,\chi \sim \exp(\frac{\Delta}{k_{B}T}).
\end{equation}

\begin{figure}
    \centering
    \epsfig{figure=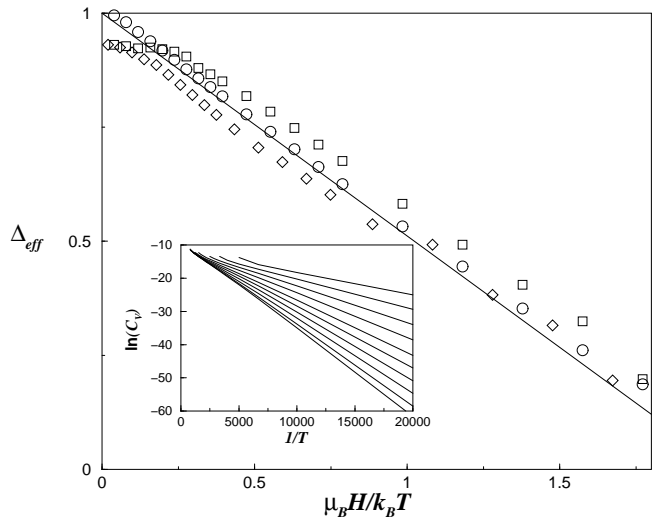, width=7cm, angle=270}
    \caption{$\Delta_{eff}$ (normalised to $\overline{|\vec{d}(\vec{k_F})|}$ at $T=H=0$)
    as a function of magnetic field parallel to $\vec{d}(\vec{k})$ extrapolated from heat capacity (circles), magnetisation
    (squares) and magnetic susceptibility
    (diamonds). The line is $\overline{|\vec{d}(\vec{k_F})|}-\mu_BH$. Inset - Logarithmic plot of heat capacity with inverse temperature at various fields. From the bottom up: H=0T, 0.28T, 0.42T, 0.71T, 0.85T, 1.13T, 1.41T, 1.76T, 2.12T, 2.47T and 2.82T.}
\label{fig:para_Deff}
\end{figure}

We find that for an anisotropic, nodeless, p-wave gap the
thermodynamics have the same form, even in the presence of a
magnetic field (see inset Fig. \ref{fig:para_Deff}) We therefore
define the effective gap, $\Delta_{eff}$ \lq seen' by the
thermodynamic functions as

\begin{equation}
C_v,M,\chi \sim \exp(\frac{\Delta_{eff}}{k_{B}T}).
\end{equation}
We find that $\Delta_{eff}$ is the mean gap at the Fermi surface,
$\overline{|\vec{d}(\vec{k_F})|}$ in zero field and that
$\Delta_{eff}$ is a linear function of magnetic field strength
(see Fig. \ref{fig:para_Deff}.) That is to say that

\begin{equation}
\Delta_{eff}=\overline{|\vec{d}(\vec{k_F})|}-\mu_B|\vec{H}|.
\end{equation}

\subsection{$\vec{d}(\vec{k})$ perpendicular to $\vec{H}$}

\begin{figure}
    \centering
    \epsfig{figure=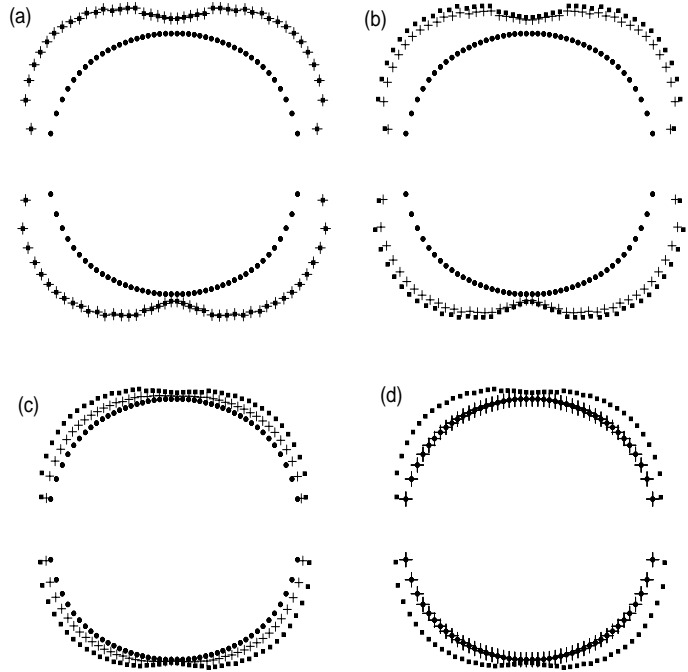, width=9cm, height=9cm, angle=270}
    \caption{The Fermi surface (circles), spin up gap (crosses) and the spin down down (squares). In the (a) A-phase $(T=H=0)$, (b) the A$_2$-phase (T=0, H=1.4T), (c) the A$_2$-phase with a larger $\kappa$ $(T=1.8K, H=1.4T)$ and (d) the A$_1$phase - not observed.}
\label{fig:perp_gap}
\end{figure}

Recall that the ground state of the model in zero field is

\begin{equation}
\vec{d}= \Delta_{0} (\sin{k_{x}}+\I\sin{k_{y}})(1,0,0).
\end{equation}
(See Fig. \ref{fig:perp_gap}a.) We will now examine the numerical
solutions of the full BdG equations corresponding to the second
solution of the Ginzburg--Landau theory (\ref{eqn:A1-phasea})
and(\ref{eqn:A1-phaseb}). In a magnetic field the ground state is
when the vector order parameter points perpendicular to the field.
There is also a change in the pairing state to a phase analogous
to the A$_2$-phase of $^3$He, where

\begin{equation}
\vec{d}= \Delta_{0} (\sin{k_{x}}+\I\sin{k_{y}})(1,\I\kappa,0).
\end{equation}
where $\kappa$ is a real function of temperature and field
strength (Fig. \ref{fig:perp_gap}b,c.) Physically this corresponds
to the majority of the spin 1 Cooper pairs aligning themselves
antiparallel to the magnetic field.

In \He as the field and temperature increase $\kappa$ increases
until $\kappa = 1$. This is the A$_1$ phase which is the ground
state of \He near to $T_C$ in finite fields. The A$_1$-phase has
order parameter

\begin{equation}
\vec{d}= \Delta_{0} (\sin{k_{x}}+\I\sin{k_{y}})(1,\I,0)
\end{equation}
and corresponds to single spin pairing with all of the Cooper
pairs aligning themselves with the magnetic field (Fig.
\ref{fig:perp_gap}d.) However, even near $T_C$ and in large fields
we do not find that the A$_1$-phase is the ground state of our
model. If such a transition does occur then it is certainly well
above the experimentally observed upper critical field. This is in
agreement with experiment as no A$_1$-phase has been observed to
date.

\begin{figure}
    \centering
    \epsfig{figure=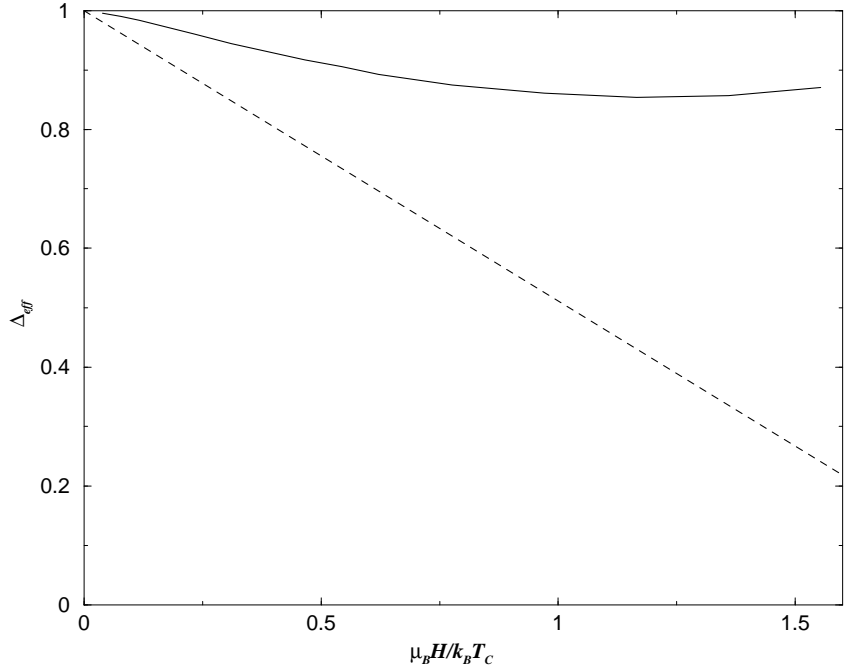, width=9cm, angle=270}
    \caption{$\Delta_{eff}$ (normalised to $\overline{|\vec{d}(\vec{k_F})|}(T=H=0)$)
    as a function of $\vec{H}$ perpendicular to $\vec{d}(\vec{k})$
    (solid line) extrapolated from heat capacity. For comparison we
    plot $\Delta_{eff}$ for $\vec{H}$ parallel to
    $\vec{d}(\vec{k})$ (dashed line).}
\label{fig:perp_Deff}
\end{figure}

Due to the nodeless gap in the A$_2$ phase the specific heat has
an exponential temperature dependence. Hence we can calculate the
effective gap for this field orientation (Fig.
\ref{fig:perp_Deff}.) We find a linear field dependence in low
fields but its dependence is much weaker than for
$\vec{d}(\vec{k})$ parallel to $\vec{H}$ and there is an upturn in
large fields. There is known to be a qualitative change in heat
capacity in this field orientation \cite{NishiZaki}. It remains to
be seen if these are related.

\section{Conclusions}

We investigated the order parameter of \Sr in an exchange-only
magnetic field. A Ginzburg--Landau symmetry analysis implied three
possibilities: either a \He A$_1$ or A$_2$ phase with
$\vec{d}(\vec{k})$ perpendicular to the magnetic field or a pure
\He A phase with $\vec{d}(\vec{k})$ parallel to the magnetic
field. We explored the exchange field dependence of the order
parameter and energy gap in a one-band model of Sr$_2$RuO$_4$. The
numerical solutions showed no A$_1$ phase for physically
reasonable field strengths and that of the two remaining phases
the A$_2$ phase is lower in free energy. We did not include the
effect of spin-orbit coupling which could change the ground state
for particular orientations of the magnetic field (particularly
with $\vec{H}$ parallel to the c-axis of the crystal.) We
investigated the behaviour of the heat capacity as a function of
both field and temperature for both of these solutions. We have
shown that the variation of the exponential cutoff below $T_C$ as
a function of $\vec{H}$ is quantitatively and qualitatively
different for these two phases. This makes heat capacity an
excellent experimental probe of the symmetry state in a magnetic
field.

We acknowledge support of BJP from an EPSRC studentship.

\section*{Appendix}{\label{Appendix:proof}}

For an A-phase triplet superconductor with $\vec{H}$ parallel to
$\vec{d}(\vec{k})$ and $\vec{z}$ the BdG equations are

\begin{eqnarray}
\begin{pmatrix}
\epsilon _{\vec{k}}+\mu_BH & 0 & 0 & d_{3}({\vec{k}})
\\ 0 &\epsilon _{\vec{k}}-\mu_BH & d_{3}({\vec{k}}) & 0
\\ 0 & d^\ast_{3}({\vec{k}}) &-\epsilon_{-\vec{k}}-\mu_BH &\ 0
\\ d^\ast_{3}({\vec{k}}) & 0 & 0 &-\epsilon_{-\vec{k}}+\mu_BH
\end{pmatrix}
\begin{pmatrix}
u_{\uparrow\sigma}({\vec{k}}) \\ u_{\downarrow\sigma}({\vec{k}}) \\
v_{\uparrow\sigma}({\vec{k}}) \\ v_{\downarrow\sigma}({\vec{k}})
\end{pmatrix}
 \notag \\ \hspace{80mm} = E_{\sigma}({\vec{k}})
\begin{pmatrix}
u_{\uparrow\sigma}({\vec{k}}) \\ u_{\downarrow\sigma}({\vec{k}}) \\
v_{\uparrow\sigma}({\vec{k}}) \\ v_{\downarrow\sigma}({\vec{k}})
\end{pmatrix}.
\end{eqnarray}
Hence, the eigenvalues are

\begin{equation}
E_\sigma=E_0(\vec{k})+\sigma\mu_BH
\end{equation}
where
\begin{equation}
E_0=\sqrt{\epsilon_{\vec{k}}+|d_3(\vec{k})|^2}
\end{equation}
is the spectrum in zero field. The eigenvectors are

\begin{equation}
u_{\sigma\sigma}(\vec{k})=\frac{d_3(\vec{k})}{\sqrt{(E_0(\vec{k})-\epsilon_{\vec{k}})^2+|d_3(\vec{k})|^2}}
\end{equation}
and
\begin{equation}
v_{\sigma-\sigma}(\vec{k})=\frac{E_0(\vec{k})-\epsilon_{\vec{k}}}{\sqrt{(E_0(\vec{k})-\epsilon_{\vec{k}})^2+|d_3(\vec{k})|^2}}.
\end{equation}

Substituting these into the self-consistency condition
(\ref{eqn:self_const}) we find that the gap equation is

\begin{equation}
d_3(\vec{k})=\frac{1}{2}\sum_{\vec{k}\sigma}U^{\sigma-\sigma}(\vec{k})\frac{d_3(\vec{k})}
{E_0(\vec{k})}\tanh(\frac{E_0(\vec{k})+\sigma\mu_BH}{2k_BT})
\end{equation}
At $T=0$ this becomes
\begin{equation}
d_3(\vec{k})=\frac{1}{2}\sum_{\vec{k}\sigma}U^{\sigma-\sigma}(\vec{k})\frac{d_3(\vec{k})}
{E_0(\vec{k})}.
\end{equation}
which is independent of $\vec{H}$.

\end{document}